\documentclass[prd,11pt,tightenlines,nofootinbib]{revtex4}
\usepackage{amsmath,amsthm,amssymb}

\begin{document}
%


\title{Comment on ``Late-time tails of a self-gravitating massless scalar field revisited'' by Bizo{\'n} et al:
The leading order asymptotics 
}

\author{Nikodem Szpak}
\affiliation{Max-Planck-Institut f\"{u}r
Gravitationsphysik, Albert-Einstein-Institut, Golm, Germany}
\date{\today}

\begin{abstract}
In \textit{Class. Quantum Grav. 26 (2009) 175006} Bizo{\'n} et al discuss the power-law tail in the long-time evolution of a spherically symmetric self-gravitating massless scalar field in odd spatial dimensions. They derive explicit expressions for the leading order asymptotics for solutions with small initial data by using formal series expansions. Unfortunately, this approach 
misses an interesting observation that the actual decay rate is a product of asymptotic cancellations occurring due to a special structure of the nonlinear terms.
Here, we show that one can 
calculate the leading asymptotics more directly by recognizing the special structure and cancellations already on the level of the wave equation.
\end{abstract}

\maketitle

\newcommand{\mybox}[2][blue]{{\color{#1}\fbox{\color{black} #2}}}
\def\blue#1{{\color{blue}#1}}
\def\d{\partial}
\def\Ocal{{\cal O}}
\def\n#1{\langle #1 \rangle}

Since the works of John \cite{John-blowup} and Asakura \cite{Asakura} who studied nonlinear wave equations with power nonlinearities it is known that the late-time asymptotics can be ruled
by the nonlinearity even for solutions starting from small initial data (for which one might naively want to ignore the nonlinear terms as causing only ``higher order'' corrections). For wave equations with nonlinearities containing first derivatives the asymptotics may even depend on the particular linear combination of the terms. Christodoulou \cite{Christodoulou-NullCond} and Klainerman \cite{Klainerman-NullCond} discovored special \textit{null structures}, which lead to a faster than generic decay or decide about a global existence of solutions. The underlying mechanism is based on asymptotic cancellations of the leading order terms in these special nonlinear structures (cf. \cite{NS-Semilin} for a detailed analysis of such cancellations).
The purpose of this Comment is to demonstrate that the same phenomenon occurs here, in the wave equation for the scalar field, and make use of it to simplify the calculation of the leading asymptotics at late-times.

In the commented Article \cite{PB_GR+Scalar}, later referred to as BCR, evolution of a self-gravitating real massless scalar field $\phi$ is considered. The Einstein equations for a $d+1$-dimensional metric with odd $d\geq 3$ restricted to spherical symmetry
\begin{equation}
ds^2 = e^{2\alpha(t,r)}\left(-e^{2\beta(t,r)} dt^2 + dr^2\right) + r^2 d \Omega_{d-1}^2\,,
\end{equation}
are analyzed, where $d\Omega_{d-1}^2$ is the round metric on the unit  $(d-1)$--dimensional sphere.
The scalar field satisfies a (quasilinear) wave equation
with smooth, and compactly supported initial data $(\phi, \dot{\phi})_{t=0} = (\varepsilon f, \varepsilon g)$ where $\varepsilon$ is a small number.
The functions $\phi, \beta$ and $m=(1-e^{-2\alpha}) r^{d-2}$ are formally expanded in the Taylor series in $\varepsilon$ about Minkowski spacetime $m_0=\beta_0=\phi_0=0$ and substituted into the field equations. This gives an infinite hierarchy of equations on $\phi_i, \beta_i, m_i$, $i=1,2,3,...$ which can be solved recursively.

At the first order $m_1=\beta_1=0$ while
\begin{align} \label{Box_phi1}
  \Box \phi_1 &= 0, & (\phi_1,\dot{\phi}_1)_{t=0} &= (f,g)
\end{align}
can be solved explicitly.
At the second order $\phi_2=0$ and
\begin{eqnarray}
\label{m2prime3+1} m'_2 &=& \kappa\, r^{d-1} \left( \dot{\phi}_1^2 + {\phi'}_1^2 \right)\,,
\\
\label{m2dot3+1} \dot{m}_2 &=&  2 \kappa\, r^{d-1} \dot{\phi}_1 \phi'_1\,,
\\
\label{beta2prime3+1} \beta'_2 &=& \frac{(d-2) m_2}{r^{d-1}}\,
\end{eqnarray}
where $\kappa=\dfrac{8 \pi}{d-1}$.
And at the third order
\begin{equation} \label{Box_phi3}
\Box \phi_3  = 2 \beta_2 \ddot{\phi}_1 + \dot{\beta}_2 \dot{\phi}_1+ \beta'_2
\phi'_1.
\end{equation}
In BCR it is claimed that the late-time asymptotics of $\phi$ is dominated by that of $\phi_3$. In order to calculate it some of the functions $\phi_i, \beta_i, m_i$ must additionally be expanded in powers of $r^{-1}$ and then inserted into \eqref{Box_phi3}. Here, cancellations of the leading terms occur since this equation has a very special structure of the nonlinear terms originating from the wave equation for $\phi$. Hence, next-to-leading order terms become important in the asymptotics.

Below, we present a method which makes use of these cancellations already at the level of the wave equation (cancellation of terms before their evaluation) and thus reduces the amount of necessary asymptotic information about the source functions on the right-hand side.
In Section \ref{sec:cancellations} we regroup the nonlinear terms, eliminate the subdominant ones and calculate the leading asymptotics for $\phi_3$ by solely evaluating the dominant nonlinear term.
In Section \ref{sec:estimates} we prove estimates which give a rigorous background for the term selection in Section \ref{sec:cancellations}.

\section{Analysis of the leading order asymptotics} \label{sec:cancellations}

\subsection{3+1 dimensions}

For $d=3$ the solution of the free wave equation \eqref{Box_phi1} can be written as (cf. (13)-(14) of BCR with $l=0$)
\begin{equation}\label{phil=0A}
    \phi_1(t,r)=\frac{a(u)-a(v)}{r}\,.
\end{equation}
where $u=t-r, v=t+r$ and the function $a$ is determined by $f$ and $g$ and has compact support in $[-R,+R]$.
The most obvious way of calculating the asymptotics of $\phi$ is to substitute the above function into \eqref{m2prime3+1}-\eqref{beta2prime3+1}, calculate $\dot{\phi}_1, \phi'_1, \ddot{\phi}_1, \phi''_1, \dot\beta_2, \beta'_2$ and insert into \eqref{Box_phi3} to obtain the desired decay in time for $\phi_3$, as was done in Section III of BCR. However, as follows from the rough estimates (cf. those obtained in Section \ref{sec:estimates}), the right-hand side of \eqref{Box_phi3} is a function supported in the vicinity of the lightcone $t=r$ and decaying like $1/r^3$. It suggests that $\phi_3$ should decay in time like $1/t^2$. As the section III of BCR shows, the true decay is by one power faster. Indeed, there happens a cancellation of leading terms in the asymptotic expansion. Here, we want to explain the cancellation mechanism already on the level of the differential equation by regrouping terms to form special structures. This transformation also allows for a considerable simplification of the calculations.

By introducing null derivatives $\d_\pm:=\frac{1}{2}(\d_t\pm\d_r)$ and rearranging terms the equation \eqref{Box_phi3} can be rewritten as
\begin{equation} \label{Box_phi3A}
\Box \phi_3  = -\frac{1}{r}\beta_2 \d_- \phi_1  + 2\d_+ \beta_2 \d_+ \phi_1 + \frac{2}{r}\d_- (r \beta_2 \d_- \phi_1) + \frac{2}{r}\beta_2 \d_+^2 (r\phi_1).
\end{equation}
The last term is identically zero for $t>R$ because there $\phi_1$ is a purely outgoing wave \eqref{phil=0A}. The second last term, in the process of inversion of $\Box = \frac{1}{r}\d_- \d_+(r\cdot)$, will turn out to be a complete derivative in the ingoing direction $u$. Since $r \beta_2 \d_- \phi_1$ has compact support in $u$ this term will vanish after integration. The second term of the above expression, as is explained in Section \ref{sec:estimates}, has faster decay in $r$ (by at least one power) than any other combination of null derivatives and together with the compact support in $u$ (localization near to the lightcone) leads to a faster decay in time for $\phi_3$. Hence, it is the first term that will determine the asymptotic behavior of $\phi_3$ at late times (see Section \ref{sec:estimates} for proof). We denote it symbolically
\begin{equation} \label{Box_phi3A-approx}
\Box \phi_3  \cong -\frac{1}{r}\beta_2 \d_- \phi_1.
\end{equation}
Hence, all we need to calculate is the leading order behavior of $\beta_2$. Substituting \eqref{phil=0A} into (\ref{m2prime3+1}) and integrating from $r=0$ we first find for $t>R$
\begin{equation}
m_2(t,r) = 4\pi \left( 2 \int \limits_{t-r}^\infty a'^2(x) \, dx - \frac {a^2(t-r)}
{r}\right) \cong 8\pi F(t-r) + \Ocal(r^{-1}),
\end{equation}
where $F(u):=I_1^0(u)$ is defined in (28) of BCR. For $t>r+R$ we have exactly $m_2(t,r)=0$.
Next, integrating equation \eqref{beta2prime3+1} from $r=0$ and using $F(u)=0$ for $u>R$ we get for $r+R>t>R$
\begin{equation}
\begin{split}
\beta_2(t,r)  \cong &
8\pi  \int \limits_{t-r}^R \frac {1} {(t-u')^2} [F(u') +\Ocal(t^{-1})] \, du'  \\
\cong & \frac{8\pi}{t^2}  \int \limits_{t-r}^\infty F(u') \, du' +\Ocal( t^{-3})
= \frac{8\pi}{t^2}  G(t-r) +\Ocal(t^{-3})
\end{split}
\end{equation}
where $G(u):=I_1^1(u)$. Otherwise, for $t>r+R$, we have $\beta_2(t,r)=0$.
Substituting this and (\ref{phil=0A}) into \eqref{Box_phi3A-approx} and using (21) of BCR (with $l=0$) we get
\begin{eqnarray}
\label{phi3(3)A} \phi_3(t,r) &=& - \frac {2^4 \pi} {r} \int \limits_{-R}^{+R} d\eta \int
\limits_{t-r}^{t+r} \frac {d\xi} {(\xi-\eta)(\xi+\eta)^2}  \left[
 G(\eta) a'(\eta)\!  + \!\mathcal{O} \left( \frac
{1} {(\xi+\eta)} \right) \right] .
\end{eqnarray}
Now, for $t\pm r\gg R$ elementary integration over $\xi$ and  by parts over $\eta$ yields the asymptotic result (34) of BCR
\begin{align}
\label{phi3tailA} \phi_3(t,r) &= \frac{t}{(t^2-r^2)^2}\left[\Gamma_0 + \mathcal{O} \left( \frac
{t} {t^2-r^2} \right)\right], & \Gamma_0 &:= - 2^5 \pi \int \limits_{-\infty}^{+\infty} F(u) a(u)\, du.
\end{align}

\subsection{Higher dimensions}

At present we cannot rigorously prove the leading order asymptotics due to a lack of an optimal decay estimate for the wave equation in d+1 dimensions, but analogously to the the 3+1 case, we are able to regroup and estimate the right-hand side terms of the wave equation for $\phi_3$ and so determine the leading order source term (see Section \ref{sec:estimates}).

We can rearrange the nonlinear terms in the equation \eqref{Box_phi3} to write them as
\begin{equation} \label{Box_phiDA}
  \Box \phi_3  = \frac{2}{r^{d-1}}\d_-\left(r^{d-1} \beta_2 \d_- \phi_1\right)
  + \frac{2}{r^{d-1}}\d_+\left(r^{d-1} \beta_2 \d_+ \phi_1\right)
\end{equation}
The second source term of the above expression, as is explained in Section \ref{sec:estimates}, has faster decay in $r$ (by at least one power) than any other combination of null derivatives and together with the compact support in $u$ leads to a faster decay in time for $\phi_3$. Hence, it is the first term that will determine the asymptotic behavior of $\phi_3$ at late times. See Section \ref{sec:estimates} for a quantitative analysis. We denote it symbolically
\begin{equation} \label{Box_phi3D-approx}
\Box \phi_3  \cong \frac{2}{r^{d-1}}\d_-\left(r^{d-1} \beta_2 \d_- \phi_1\right) .
\end{equation}
Using (21) of BCR we get
\begin{eqnarray}
\phi_3(t,r) &=& \frac {1} {2^{l+2}r^{l+1}} \int_{t-r}^{t+r} d\xi \int_{-\infty}^{+\infty} d\eta\; \frac{P_l(\mu)}{(\xi-\eta)^{l+1}} \d_\eta \left[(\xi-\eta)^{2l+2} \beta_2 \d_\eta \phi_1\right]
\end{eqnarray}
where $l=(d-3)/2$.
The inner integral over $\eta$ can be integrated by parts (it produces no boundary terms since the integrand has compact support in $\eta$)
\begin{eqnarray} \label{phi3(d+1)A}
\phi_3(t,r) &=& -\frac {1} {2^{l+2}r^{l+1}} \int_{t-r}^{t+r} d\xi \int_{-\infty}^{+\infty} d\eta\; \d_\eta\left[\frac{P_l(\mu)}{(\xi-\eta)^{l+1}}\right] (\xi-\eta)^{2l+2} \beta_2 \d_\eta \phi_1.
\end{eqnarray}
Now we only need to find the asymptotic form of the expression $\beta_2 \d_- \phi_1$. According to (13)-(14) of BCR we have
\begin{align}\label{phil=0B}
    \phi_1(t,r)&\cong\frac{a^{(l)}(u)}{r^{1+l}}\,+\Ocal(r^{-2-l}),&
    \dot\phi_1(t,r)&\cong\frac{a^{(l+1)}(u)}{r^{1+l}}\,+\Ocal(r^{-2-l}),&
    \phi'_1(t,r)&\cong -\frac{a^{(l+1)}(u)}{r^{1+l}}\,+\Ocal(r^{-2-l}),&
\end{align}
and
\begin{equation}
m_2(t,r) \stackrel{t>R}{\cong} 2\kappa \int \limits_{t-r}^\infty [a^{(l+1)}(x)]^2 \, dx + \Ocal(r^{-1}) =: 2\kappa F_l(t-r) + \Ocal(r^{-1}),
\end{equation}
where $F_l(u):=I_{l+1}^1(u)$. Again, for $t>r+R$ we have exactly $m_2(t,r)=0$.
Next, integrating equation \eqref{beta2prime3+1} from $r=0$ and using $F_l(u)=0$ for $u>R$ we get for $r+R>t>R$
\begin{equation}
\begin{split}
\beta_2(t,r)  \cong &
(d-2)2\kappa  \int \limits_{t-r}^R \frac {1} {(t-u')^{2+2l}} [F_l(u') +\Ocal(t^{-1})] \, du'  \\
\cong & \frac{(2l+1)}{(2l+2)}\frac{16\pi}{t^{2+2l}}  \int \limits_{u}^\infty F_l(u') \, du' +\Ocal(t^{-3-2l}) +\Ocal( t^{-3-2l})
= \frac{(2l+1)}{(2l+2)}\frac{16\pi}{t^{2+2l}}  G_l(u) +\Ocal(t^{-3-2l})
\end{split}
\end{equation}
where $G_l(u):=I_{l+1}^1(u)$, otherwise, for $t>r+R$, we have $\beta_2(t,r)=0$.
Substituting this and (\ref{phil=0B}) into \eqref{phi3(d+1)A} we obtain
\begin{eqnarray} \label{phi3(d+1)A'}
\phi_3(t,r) &=& -\frac{(2l+1)}{(2l+2)}\frac{2^{2l+5} \pi}{r^{l+1}} \int_{t-r}^{t+r} d\xi \int_{-\infty}^{+\infty} d\eta\; \d_\eta\left[\frac{P_l(\mu)}{(\xi-\eta)^{l+1}}\right]  \frac{(\xi-\eta)^{l+1}}{(\xi+\eta)^{2+2l}} G_l(\eta) a^{(l+1)}(\eta).
\end{eqnarray}
Analogously to (22) or BCR it holds
\begin{equation} \label{int-P}
  \int \limits_{t-r}^{t+r} d\xi \, \d_\eta\left[\frac{P_l(\mu)}{(\xi-\eta)^{l+1}}\right] \frac{(\xi-\eta)^{l+1}}{(\xi+\eta)^{2+2l}} \cong
  (-1)^l 2^l \frac{(2l+2)}{(2l+1)} \, \frac {r^{l+1}} {t^{3+3l}} \left[ 1
  + \mathcal{O} \left( \frac {1} {t} \right) \right],
\end{equation}
which applied to the above integral yields the asymptotic result (44) of BCR
\begin{align} \label{phi3tail(d+1)A}
\phi_3(t,r) &= \frac{\Gamma_l}{t^{3l+3}} + \mathcal{O} \left( \frac{1} {t^{3l+4}} \right),&
\Gamma_l &:= (-1)^{l+1} 2^{3l+5} \pi \int \limits_{-\infty}^{+\infty} F_l(\eta) a^{(l)}(\eta)\, d\eta,
\end{align}
for late times ($t\gg r$), where in $\Gamma_l$ we integrated by parts over $\eta$.
The asymptotic formula (45) of BCR can be straightforwardly derived from \eqref{phi3(d+1)A'} as well, it only requires some other asymptotic expansion of the integral \eqref{int-P}.

\section{Estimates} \label{sec:estimates}

\subsection{3+1 dimensions}

In this section we will prove decay estimates for $m_2, \beta_2$ and $\phi_3$. They will lose information about the exact amplitudes but will be helpful in separating the leading asymptotics from the subleading corrections decaying faster.

From \eqref{phil=0A} we know that $\phi_1$ is supported in the strip $0<t-R\leq r\leq t+R$ and can estimate there its derivatives
\begin{align}
  |\phi_1(t,r)|, |\dot\phi_1(t,r)|, |\phi'_1(t,r)| &\lesssim \frac{1}{\n{r}}, &
  | \dot{\phi}_1 + {\phi'_1} | &\lesssim \frac{1}{\n{r}^2},
\end{align}
where $\n{x}:=1+|x|$ and ``$\lesssim$'' means ``less or equal than'' up to some multiplicative constant which we skip for brevity. Observe that for functions supported in the strip $|t-r|<R$ estimates by powers of $\n{r}$ and $\n{t}$ are equivalent, since there exist constants $C_1, C_2$ such that $C_1 \n{t} \leq \n{r} \leq C_2 \n{t}$.

From \eqref{m2prime3+1} we have $m_2(t,r)=0$ for $r<t-R$ and for $0<t-R\leq r\leq t+R$ we estimate
\begin{equation}
  |m_2(t,r)| \leq \int_0^r {r'}^2 \left( |\dot{\phi}_1|^2 + |\phi'_1|^2 \right)\; dr' \lesssim \int_{t-R}^{r} \frac{r'^2}{\n{r'}^2} dr' \leq C
\end{equation}
where $C$ depends only on $R$. Actually, there is a universal bound $|m_2(t,r)|\leq M$ where $M$ is the total energy (``ADM mass'') of $\phi_1$. Moreover, for small $r$, say $r<1$, we immediately see that $|m_2(t,r)| \lesssim r^3$. These two facts can be put together to give
\begin{equation}
  |m_2(t,r)| \lesssim   \frac{r^3}{\n{r}^3}
\end{equation}
Next, from \eqref{beta2prime3+1} we get $\beta_2(t,r)=0$ for $r<t-R$ and for $0<t-R\leq r\leq t+R$ we again estimate
\begin{equation}
  |\beta_2(t,r)| \leq \int_0^r \frac{m_2(t,r')}{{r'}^2}\; dr' \lesssim \int_{t-R}^{t+R} \frac{1}{\n{r'}^2} dr' \leq \frac{2R}{\n{t-R}\n{t+R}} \lesssim \frac{1}{\n{t}^2}
\end{equation}
We will also need a similar estimate for the outgoing derivative of $\beta_2$. Therefore we combine the equations \eqref{m2prime3+1}-\eqref{beta2prime3+1} to get the identity
\begin{equation}
  \d_+ \beta_2 = \dot\beta_2 + \beta'_2 = \int_0^r \left( \dot{\phi}_1 + {\phi'_1} \right)^2 dr' - 2 \int_0^r \frac{m_2}{{r'}^3} dr'.
\end{equation}
Using the above estimates on $\dot{\phi}_1 + {\phi'_1}$ and $m_2$ we find
\begin{equation}
  |\d_+ \beta_2| \lesssim \int_{t-R}^{t+R} \frac{1}{\n{r'}^4} dr' + \int_{t-R}^{t+R} \frac{1}{\n{r'}^3} dr' \lesssim \frac{1}{\n{t}^3}
\end{equation}
for $0<t-R\leq r\leq t+R$ as well as $\d_+ \beta_2(t,r)=0$ for $r<t-R$.

Finally, we analyze the various source terms in the wave equation \eqref{Box_phi3}. Let us split the solution into four components introduced in \eqref{Box_phi3A}
\begin{align}
  \Box \phi_{3A} &= -\frac{1}{r}\beta_2 \d_- \phi_1, &
  \Box \phi_{3B} &= 2\d_+ \beta_2 \d_+ \phi_1, \\
  \Box \phi_{3C} &= \frac{2}{r}\d_- (r \beta_2 \d_- \phi_1), &
  \Box \phi_{3D} &= \frac{2}{r}\beta_2 \d_+^2 (r\phi_1),
\end{align}
with $\phi_3=\phi_{3A}+\phi_{3B}+\phi_{3C}+\phi_{3D}$. All terms on the right hand side are supported in $|t-r|<R$. By the above bounds we can estimate the first two components
\begin{align}
  \left|\frac{1}{r}\beta_2 \d_- \phi_1\right| &\lesssim \frac{1}{r\n{r}\n{t}^2} \lesssim \frac{1}{r\n{r}^3}, &
  \left|\d_+ \beta_2 \d_+ \phi_1\right| &\lesssim \frac{1}{\n{t}^3\n{r}^2} \lesssim \frac{1}{\n{r}^5}
\end{align}
Now, e.g. from \cite{NS-DecayLemma}, we find
\begin{align}
  |\phi_{3A}(t,r)| &\lesssim \frac{1}{\n{t+r}\n{t-r}^2}, &
  |\phi_{3B}(t,r)| &\lesssim \frac{1}{\n{t+r}\n{t-r}^3},
\end{align}
so we see that $\phi_{3B}$ becomes subdominant to $\phi_{3A}$ regarding the late time asymptotics.

Next, in spherical symmetry $\Box \phi_{3C} \equiv \frac{4}{r}\d_- \d_+(r \phi_{3C})$ and hence
\begin{equation}
  \phi_{3C}(t,r) = \frac{1}{4r} \int_{t-r}^{t+r} du \int_{-(t+r)}^{t-r} dv\; \d_v (r \beta_2 \d_v \phi_1).
\end{equation}
The inner integral is an integral of a total derivative of a compactly supported function of $v$. The integration range is bigger than its support $[-R,R]$ so the integral vanishes. This gives $\phi_{3C}=0$ for $t>r+R$.

The last component $\phi_{3D}=0$ for $t>r+R$ because the source vanishes identically in the integration region when inverting the wave operator as we did above.

Finally, for late times, in the region $t>r+R$,
\begin{equation}
  |\phi_{3}(t,r)| \cong |\phi_{3A}(t,r)| \lesssim \frac{1}{\n{t+r}\n{t-r}^2}
\end{equation}
and other components of $\phi$ can be a priori estimated to be subdominant. Therefore, in the asymptotic analysis, concerned solely with the leading order behavior, it is sufficient to keep only the first source term in \eqref{Box_phi3A}.

\subsection{Higher dimensions}

Here, we want to show that the second source term in \eqref{Box_phiDA} is asymptotically subdominant with respect to the first one.
In dimension d+1, the estimates for $\phi_1, m_2$ and $\beta_2$ and their derivatives can be obtained analogously to the 3+1 case and read for $0<t-R\leq r\leq t+R$
\begin{align}
  |\phi_1(t,r)|, |\dot\phi_1(t,r)|, |\phi'_1(t,r)| &\lesssim \frac{1}{\n{r}^{1+l}}, &
  | \d_+^k \phi_1| &\lesssim \frac{1}{\n{r}^{1+l+k}},&
  | \d_-^k \phi_1| &\lesssim \frac{1}{\n{r}^{1+l}},
\end{align}
\begin{align}
  |m_2(t,r)| &\lesssim  C \frac{r^d}{\n{r}^d},&
  |\beta_2(t,r)| &\lesssim \frac{1}{\n{t}^{d-1}}, &
  |\d_+ \beta_2| &\lesssim \frac{1}{\n{t}^d},&
  |\d_- \beta_2| &\lesssim \frac{1}{\n{t}^{d-1}}
\end{align}
while all these functions vanish for $r<t-R$. It allows us to control both source terms in \eqref{Box_phiDA}
\begin{align}
  \left|\frac{1}{r^{d-1}}\d_-\left(r^{d-1} \beta_2 \d_- \phi_1\right) \right| &\leq
  \left|\d_- \beta_2 \d_- \phi_1\right| +
  \left|\beta_2 \d_-^2 \phi_1\right| +
  \left|\frac{(d-1)}{r}\beta_2 \d_- \phi_1\right| \\ \nonumber
  &\lesssim \frac{1}{\n{t}^{d-1}\n{r}^{1+l}} + \frac{1}{\n{t}^{d-1}\n{r}^{1+l}}.+ \frac{1}{r\n{t}^{d-1}\n{r}^{1+l}}
  \lesssim \frac{1}{\n{r}^{3+3l}}\\
  \left|\frac{1}{r^{d-1}}\d_+\left(r^{d-1} \beta_2 \d_+ \phi_1\right) \right| &\leq
  \left|\d_+ \beta_2 \d_+ \phi_1\right| +
  \left|\beta_2 \d_+^2 \phi_1\right| +
  \left|\frac{(d-1)}{r}\beta_2 \d_+ \phi_1\right|\\ \nonumber
  &\lesssim \frac{1}{\n{t}^{d}\n{r}^{2+l}} + \frac{1}{\n{t}^{d-1}\n{r}^{3+l}}.+ \frac{1}{r\n{t}^{d-1}\n{r}^{2+l}}
  \lesssim \frac{1}{\n{r}^{5+3l}}.
\end{align}
Both estimates are optimal (c.f. the explicit asymptotic expressions in Section IV of BCR), hence the second therm is indeed subdominant what justifies neglecting it in the leading order calculations.

However, there is a problem in d+1 dimensions which is absent in 3+1. We lack an optimal decay estimate for the wave equation in d+1 dimensions, given a source with prescribed decay. The presently best known estimates (c.f. \cite{Karageorgis,Kubo-SemilinDecay}) lose $l$ powers in the late-time decay relative to what is optimal. Therefore, for the above sources we are able to show rigorously only the decay
\begin{equation}
  |\phi_{3}(t,r)| \lesssim \frac{1}{\n{t+r}\n{t-r}^{2+2l}},
\end{equation}
while $1/t^{3+3l}$ is optimal for late times. However, it does not change the fact that in the asymptotic analysis it is the first source term in \eqref{Box_phiDA} that dominates the asymptotics of $\phi_3$ at late times.

\bibliography{QNMs}

\begin{thebibliography}{1}

\bibitem{John-blowup}
F.~John.
\newblock Blow-up of solutions of nonlinear wave equations in three space
  dimensions.
\newblock {\em Manuscripta Math.}, (28):235--268, 1979.

\bibitem{Asakura}
Asakura F.
\newblock Existence of a global solution to a semi-linear wave equation with
  slowly decreasing initial data in three space dimenstions.
\newblock {\em Comm. Part. Diff. Eq.}, 13(11):1459--1487, 1986.

\bibitem{Christodoulou-NullCond}
D.~Christodoulou.
\newblock Global solutions of nonlinear hyperbolic equations for small initial
  data.
\newblock {\em Comm. Pure Appl. Math.}, 39:267--283, 1986.

\bibitem{Klainerman-NullCond}
S.~Klainerman.
\newblock The null condition and global existence to nonlinear wave equations.
\newblock {\em Lect. Appl. Math.}, 23:293--326, 1986.

\bibitem{NS-Semilin}
N.~Szpak.
\newblock Pointwise decay estimates for semilinear wave equations containing
  first derivatives.
\newblock 2009.
\newblock in preparation.

\bibitem{PB_GR+Scalar}
P.~Bizon, T.~Chmaj, and A.~Rostworowski.
\newblock Late-time tails of a self-gravitating scalar field revisited.
\newblock {\em Class. Quantum Grav.}, 26:175006, 2009.
\newblock arXiv: gr-qc/0812.4333v3.

\bibitem{NS-DecayLemma}
N.~Szpak.
\newblock Simple proof of a useful pointwise estimate for the wave equation.
\newblock 2007.
\newblock arXiv: math-ph/0708.2801.

\bibitem{Karageorgis}
P.~Karageorgis.
\newblock Small-data scattering for nonlinear waves with potential and initial
  data of critical decay.
\newblock 2005.
\newblock arXiv: math.AP/0503208.

\bibitem{Kubo-SemilinDecay}
H.~Kubo.
\newblock Slowly decaying solutions for semilinear wave equations in odd space
  dimensions.
\newblock {\em Nonlinear Analysis, Theory, Methods \& Applications},
  28(2):327--357, 1997.

\end{thebibliography}
\bibliographystyle{unsrt}

\end{document}